\documentclass[floatfix,aps,twocolumn,showpacs,nofootinbib,superscriptaddress]{revtex4}

\usepackage{graphicx}
\usepackage{subfigure}
\usepackage{epsfig}
\usepackage{amsmath}\usepackage{amsfonts}
\usepackage{amssymb}
\usepackage{hyperref}
\usepackage{dsfont}

\newcommand{\mean}[1]{\langle #1 \rangle}
\newcommand{\ket}[1]{|#1 \rangle}
\newcommand{\cale}{\mathcal{E}}
\newcommand{\beq}{\begin{equation}}
\newcommand{\eeq}{\end{equation}}
\newcommand{\cals}{\mathcal{S}}

\begin{document}
\author{Nahuel Freitas}
\affiliation{Departamento de F\'\i sica, FCEyN, UBA, Pabell\'on 1,
Ciudad Universitaria, 1428 Buenos Aires, Argentina}
\affiliation{Instituto de F\'\i sica de Buenos Aires, UBA CONICET,
Pabell\'on 1, Ciudad Universitaria, 1428 Buenos Aires, Argentina}
\affiliation{Theoretische Physik, Universit\"at des Saarlandes, D-66123 Saarbr\"ucken, Germany}

\author{Juan Pablo Paz}
\affiliation{Departamento de F\'\i sica, FCEyN, UBA, Pabell\'on 1,
Ciudad Universitaria, 1428 Buenos Aires, Argentina}
\affiliation{Instituto de F\'\i sica de Buenos Aires, UBA CONICET,
Pabell\'on 1, Ciudad Universitaria, 1428 Buenos Aires, Argentina}

\title{How much can we cool a
quantum oscillator? \\ 
A useful analogy to understand laser
cooling as a thermodynamical process
}

\date{\today}

\begin{abstract}
We analyze the lowest achievable temperature for
a mechanical oscillator (representing, for example,
the motion of a single trapped ion) which is coupled
with a driven quantum
refrigerator. The refrigerator is composed of
a parametrically driven system
(which we also consider to
be a single oscillator in the simplest case) which is
coupled to a reservoir where the energy is dumped.
We show that the cooling of the oscillator
(that can be achieved due to the resonant
transport of its phonon excitations into
the environment)
is always stopped by a fundamental heating
process that is always dominant at sufficiently
low temperatures. This process can be
described as the non resonant production of excitation
pairs. This result is in close
analogy with the recent study that
showed that pair production
is responsible
for enforcing the validity of the dynamical version of
the third law of thermodynamics
(Phys. Rev. E 95, 012146). Interestingly, we
relate our model to
the usual ones used to describe laser cooling of a
single trapped ion and reobtaining the correct limiting
temperatures for the limits of resolved and non-resolved
sidebands. Our findings (that also serve to
estimate the lowest temperatures that can be
achieved in a variety of other situations) indicate
that the limit
for laser cooling can also be associated
with non resonant pair production. In fact,
as we show, this is the case: The limiting
temperature for laser cooling is achieved when
the cooling transitions induced
by the resonant transport of excitations from
the motion into the electromagnetic
environment is compensated
by the heating transitions
induced by the creation of
phonon-photon pairs.
\end{abstract}

\maketitle

\section{Introduction}
Cooling is a fundamental task to achieve
precise control of individual quantum
systems, and thus crucial in the development
of quantum technologies.
Atoms, optical cavities, mechanical resonators, and
electronic systems must often be cooled to
ultra low temperatures in order to avoid
spurious thermal excitations and achieve the
desired degree of control
Trapped ions are cooled down close to
their motional ground
state as the first step of any quantum
algorithm
\cite{eschner2003,oconell2010,teufel2011,vuletic1998}.
 In this context, it is important
 to understand
what are the lowest temperatures that could be
achieved by the different cooling schemes that
one can use.

The study of the
ultimate limit for cooling is,
therefore, not only an interesting fundamental
question, but
also a question which is important from a practical
point of view. This question has been recently
addressed in the context of the emerging
field known as ``Quantum Thermodynamics'' \cite{allahverdyan2011,levy2012,wu2013,ticozzi2014,masanes2017,wilming2017}

\begin{figure}[]
    \centering
    \includegraphics[scale=.9]{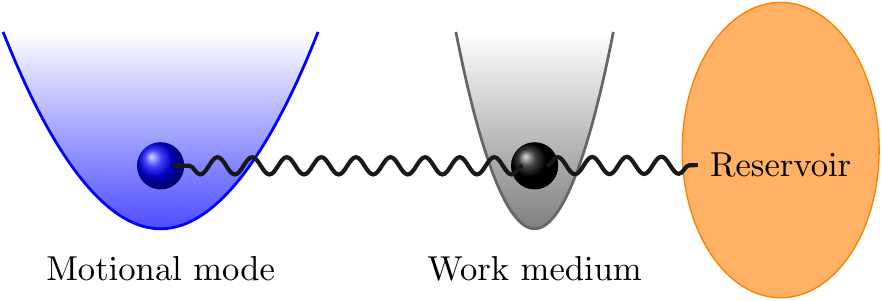}
    \caption{The model. A motional mode is
    cooled by coupling it to a
    refrigerator consisting of a driven
    harmonic oscillator, which acts as a work medium
    which is externally driven. This oscillator is itself
    coupled with
    a bosonic reservoir, which is where all
    entropy is finally dumped.}
    \label{fig:model}
\end{figure}

From a fundamental perspective, an
effort has been made to derive the third law
of thermodynamics from first principles. In fact,
the impossibility to achieve perfect ground state
cooling in
finite time (Nernst unattainability principle)
has been formally demonstrated
However, these
proofs, are not
of practical significance (as the lower bounds
in achievable temperatures are too low indeed)
and do not necessarily
offer a  direct insight on what
might be the physical processes imposing
the limitations for cooling
in actual implementations
\cite{ticozzi2014,wu2013,masanes2017,wilming2017}.

In this paper we focus on the analysis of the ultimate
cooling limit of a class of refrigerators which are
externally driven (a situation which is the most
important one in practical realizations).
We recently  presented a detailed analysis of
the behavior of a class of quantum refrigerators
\cite{freitas2017}
(linear, externally driven fridges) for which
it is possible to identify a very general process
as the one responsible for imposing the fundamental
limitation for cooling: Such process is nothing
but pair creation (a process which is relevant,
as discussed below, in other areas of physics
and is closely related with the
main mechanism behind the so called
dynamical Casimir effect, DCE
\cite{benenti2015,freitas2017}).
In fact, our work analyzed a
family of linear machines consisting of arbitrary
harmonic networks which are parametrically driven
while connected with an arbitrary number
of bosonic reservoirs, prepared at
arbitrary temperatures and characterized by
generic spectral densities (see below).
The exact solution of this model (obtained
without using common approximations
 such as the weak coupling limit between the system and its
environments, the rotating-wave approximation
or the
Markovian assumption)
was used to obtain
expressions for the dissipated work
and the heat currents flowing from each
reservoir into the system. A close examination
of this exact solution illuminates the fact that
the creation of excitation pairs is dominant
at sufficiently low temperatures and is the
process enforcing the third law. Interestingly,
in our work it was shown that an analysis
based on the
use of master equations derived in the
weak coupling limit would fail to incorporate
this fundamental process (because of the fact that
the non resonant pair creation processes are
not of leading order in the coupling between
the system and its environments ).
It was also shown that by neglecting this
fundamental term, which induces heating of
all reservoirs at sufficiently low temperatures,
the third law could indeed be
violated, as suggested and discussed in Ref.
\cite{kolavr2012}.

Here, we take advantage of the analysis
presented in Ref. \cite{freitas2017} and
apply it to analyze the limit for cooling a single
oscillator (this is possible since, as we
mentioned, the results contained in Ref.
\cite{freitas2017} are valid for environments
with arbitrary spectral densities). Analyzing the
cooling limit for a single oscillator is
relevant in several contexts, such as in the
case of cold trapped ions
\cite{diedrich1989}, trapped atoms
\cite{hamann1998}, or
micromechanical oscillators
\cite{teufel2011}. The very simple model we will
analyze and solve can be naturally viewed as
a thermal machine, or a thermal refrigerator (see Fig. \ref{fig:model}).
The machine is composed of a working medium
(a parametrically driven harmonic oscillator)
that is in simultaneous contact with two reservoirs.
One of these reservoirs has a single harmonic
mode that we want to cool. The
other reservoir (typically representing
the electromagnetic field) is where the
energy is dumped.
As we will see, the model presented here
is an interesting analogy to other
more realistic models for laser cooling.
Notably, this simple model is sufficient to
derive the lowest achievable temperatures in
the most relevant physical regimes (and to predict
their values in other, still unexplored, regimes).
Moreover, it enables a simple interpretation of the
heating process limiting laser cooling as the
non resonant creation of phonon-photon pairs.
The model also enables us to conclude that in
any laser cooling process there should be
a photon pair production process associated
with it, which is closely analogous to the pair
production process present in the DCE.
In fact, as we show below, our analysis will
enable us to estimate the rate of photon pair
production that is associated to the
typical laser cooling mechanisms with
optical frequencies and analyze
the possibility of detecting them.

The paper is organized as follows: In Section 2
we present our model to understand laser cooling
as a thermodynamical process. We solve the
long time dynamics of the model and study its
thermodynamical properties. In Section 3 we apply
this model to study the cooling of a single motional
degree of freedom. We compute the heat currents
in this case and find the lowest temperatures that
can be achieved. Not only we analyze the two
most relevant regimes (Doppler cooling and sideband
resolved cooling) but also present possible
generalizations where much lower temperatures
may be accessible. In Section 4 we analyze the
nature of the spectrum of excitations that are
created in the photonic reservoir during laser
cooling. We show that the photonic spectrum
not only consists of two peaks (corresponding
to photons emitted during cooling and heating
transitions) but also has a broad contribution
arising from photon pairs created directly from
the driving by a process which is closely connected
with the DCE. We finally summarize our results in
Section 5.

\section{A quantum refrigerator as a model
for laser cooling}

We will present here a simple model that will
enable us to study the lowest temperature
that can be achieved by laser cooling. In such
case, an atom (typically, a two level atom) is
illuminated with a laser and three types of
degrees of freedom are coupled between
each other. The internal (electronic) levels
of the atom (a system that we will denote as
$\cals$ below) couple with the quantized
electromagnetic
field (whose modes act as an environment
which we denote as $\cale_B$ below). The
internal modes also couple with the motional
degrees of freedom of the atom, which we
consider as an harmonic oscillator and
denote as $\cale_A$ below. The coupling to
the laser induces transitions between the
atomic levels. Such levels effectively
act as a pump (which is driven by the power
of the laser) that takes energy out of $\cale_A$
and dumps it into $\cale_B$. We will
describe a very simple model
to study this situation, which has the
virtue of being exactly solvable. The main
simplification is to replace the internal electronic
levels of the atom by a single harmonic mode.
Of course, this is a rough approximation which
will only be reasonable at sufficiently low
temperatures, where only the lowest energy
levels of the spectrum of
$\cals$ will matter. As we will see,
the model, whose essential ingredients are
shown in Figure 1, is able to accurately
predict the
correct limiting temperatures obtained for
laser cooling in the resolved sideband limit and
also in the non-resolved sideband case
(and it can also be used to predict new and
interesting phenomena).

\subsection{The model: Dynamics}
Here, we will describe the model and the main
properties of its solution. The rigorous derivation
of the main equations can be found in Ref.
\cite{freitas2017}. We
consider a parametrically driven
oscillator with a Hamiltonian
\beq
H_\cals= \frac{p^2}{2M} + \frac{1}{2} M V(t) x^2.
\eeq
This system couples with two others, which
we arbitrarily treat as `reservoirs' and
denote them as
$\cale_A$ and $\cale_B$. We warn
the reader that while in this
subsection
we will consider the reservoirs to
be rather general, in the following
section $\cale_A$ will
be assumed to consist of a single
oscillator while $\cale_B$ will consist
of an infinite set of bosonic
modes (the modes of the electromagnetic
field).
So, for the moment we consider the
reservoirs as
 consisting of collections
of independent oscillators
with Hamiltonians
\beq
H_\alpha = \sum_{j} \left(\frac{\pi_{\alpha,j}^2}{2m_j}
+ \frac{1}{2} {m_j\omega_{\alpha,j}^2} q_{\alpha,j}^2\right),
\eeq
where $\alpha=A,B$. The interaction between
$\cals$ and $\cale_\alpha$
is considered to be linear in both the coordinates
of the system and the environments and
is described by  the Hamiltonians
\beq
H_{\cals,\alpha} = x\: \sum_{k}
c_{\alpha,k}\:q_{\alpha,k},
\eeq
where $c_{\alpha,k}$ are the corresponding
coupling constants.
Initially, the reservoirs
are uncorrelated with $\cals$
and are prepared in thermal states at temperatures
$T_\alpha$.

We assume that the driving is
periodic and write
\beq
V(t)=\sum_k V_k e^{ik\omega_d t}.
\eeq
We also assume that the coupling
with the reservoirs induces
a stable stationary
regime in the asymptotic limit (the
long time limit).
In this regime, the state of
the system is also periodic and has
the same period of the driving, i.e.
$T=2\pi/\omega_d$).

In what follows,
the Green's function of the system
$\cals$  will be the
essential tool we will use to
analyze the
asymptotic state (which, being a Gaussian
state, is fully characterized by its covariance
matrix). The Green function is nothing but
the response
function of $\cals$
defined as the solution of the
equation
\beq
\ddot G(t,t')+V(t)G(t,t')+
\int_0^t d\tau \gamma(t-\tau)\dot G(\tau,t)=\delta(t,t'),
\eeq
where the dot denotes a derivative
with respect to the first temporal argument
and $\gamma(t)$ is the dissipation kernel which
incorporates all the effect
of the environments on the evolution of
the system. This kernel is defined as
\beq
\gamma(t)=\int_0^\infty d\omega \frac{I(\omega)}{\omega}
\cos(\omega t),
\eeq
in terms of the so called spectral density
of the environment, which is such that
$I(\omega)=\sum_\alpha
I_\alpha(\omega)$, with
\beq
I_\alpha(\omega)=
\sum_l \frac{c_{\alpha,l}^2}{M m_l\omega_l}.
\eeq
Below, we will use the Laplace
transform of $\gamma(t)$, which we
 denote as $\hat \gamma(s)$) and
turns out to be
\beq
\hat\gamma(s)=\int_0^\infty d\omega \frac{I(\omega)}{\omega}\frac{s}{s^2+\omega^2}.
\eeq

As a consequence of the driving, which
breaks time homogeneity, the Green function
$G(t,t')$ does not depend on the time
difference $(t-t')$ only. In spite of this
complication, it is possible to find
following relatively simple expression,
which is valid in the asymptotic limit
\cite{freitas2017}:
\begin{equation}
G(t,t')=\frac{1}{2\pi}
\sum_k\int_{-\infty}^\infty
d\omega A_k(\omega) e^{i\omega
(t-t')}\, e^{ik\omega_d t}.\nonumber
\end{equation}
As explained in \cite{freitas2017},
the derivation of this equation (that we omit here)
involves the use of Floquet theory for periodically
driven systems. The interpretation of  this equation is simple:
In the absence of driving, the only
surviving term in the Floquet summation
is the one with $k=0$.  In
this case, the Green function depends only
on the time difference and
$A_0(\omega)$ is simply given as
$A_0(\omega)=g(i\omega)$, where
$g(s)$ is the Laplace
transform of the undriven damped oscillator,
(see below for an explicit formula).
When the driving breaks time
homogeneity, terms with $k\neq 0$ appear
in the Floquet expansion for $G(t,t')$
giving rise to a non trivial
 dependence on the two temporal
 arguments. The Floquet
coefficients $A_k(\omega)$
are coupled between
each other and satisfy the equations
\begin{equation}
    A_k(\omega) =
  \delta_{k,0} \; g\left(i\omega\right)
  -\sum_{j \neq 0 }
  g\left(i(\omega +k \omega_d )\right)
  V_j A_{k-j}(\omega) .
  \label{meq:rel_Ak}
\end{equation}
In all above equations, the
Laplace transform of the static Green's
function can be written as
\beq
g(i\omega)= (-\omega^2 + V_R +
i\omega  \hat \gamma(i\omega))^{-1},
\eeq
which can also be rewritten as
\beq
g(i\omega)=
(\omega_0^2-(\omega-i\hat\gamma(i\omega)/2)^2)^{-1}.
\eeq
Here, $\omega_0^2=V_R-\hat\gamma^2(i\omega)/4$
and $V_R$ is the renormalized frequency of
$\cals$, which is defined as
$V_R=V_0-\gamma(0)$. The previous
definitions clearly imply
that both $\omega_0$ and
$\hat\gamma(i\omega)$
do  not depend on $\omega$
only when the total spectral density is ohmic,
i.e. when $I(\omega)
\propto\omega$. In that case,
the Green function
oscillates with frequency $\omega_0$ and
decays
with a rate $\hat\gamma\equiv\gamma$.

\subsection{Thermodynamics: Heat currents}
To study the exchange of energy between
$\cals$, $\cale_A$ and $\cale_B$ in the
stationary regime we first analyze the variation
of the expectation value of $H_\cals$ which
satisfies
\beq
\frac{d\mean{H_\cals}}{dt}=
\mean{\partial_tH_\cals}
-i\sum_\alpha\mean{[H_\cals,H_{\cals,\alpha}]},
\eeq
(we use $\hbar = 1$ here and below).
This equation enables us to identify
the notions of work and heat which are
the two
sources for the change in the energy. Thus,
the variation of the energy
induced by the explicit time dependence
of the system's Hamiltonian
is associated with
work (more precisely, with power), as
\beq
\dot{\mathcal W}=
\mean{\partial_tH_\cals/\partial t}.
\eeq
In turn,
the variation of the energy of $\cals$ arising
from the interaction with each reservoir
$\cale_\alpha$ is
associated with the heat flowing into
the system per unit time, which we
denote as $\dot
{\mathcal Q}_\alpha$ and turns out
to be
\beq
\dot{\mathcal Q}_\alpha = -i
\mean{[H_\cals,H_{\cals,\alpha}]}.
\eeq
Therefore, the above equation
is nothing but the
first law of thermodynamics, i.e.
$d\mean{H_\cals}/dt =
\dot {\mathcal W}+
\sum_\alpha
\dot{\mathcal Q}_\alpha$. In what follows we will
study the average values of the work and
the heat
currents over a driving period (in the limit
of long times). These quantities will be
respectively defined as $\dot W$ and
$\dot Q_\alpha$. In this asymptotic
regime, it can be shown that the
average heat current $\dot Q_\alpha$
coincides with the time derivative of the
energy of $\cale_\alpha$ (i.e., $\dot Q_\alpha=
-\dot{\mean{H_\alpha}}$: the energy lost
by $\cale_\alpha$ is gained by $\cals$ over
a driving period) \cite{freitas2017}.
Then, the averaged version of the
first law is simply the identity
$0=\dot W+\sum_\alpha \dot Q_\alpha$.

\begin{figure}[ht]
    \centering
    \includegraphics[width=.4\textwidth]{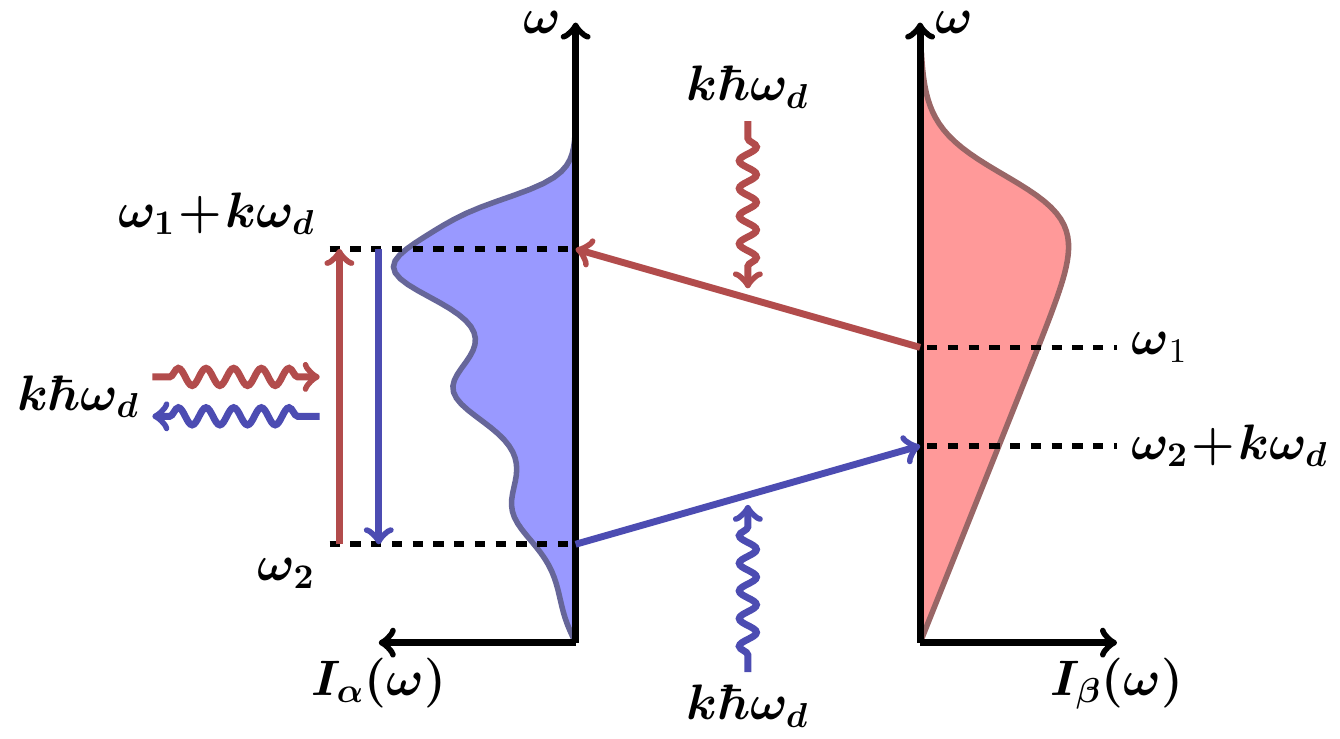}
    \caption{Resonant processes affecting the heat
    flow in or out of a thermal reservoir are of two
    types: resonant pumping (RP), associated with
    $\dot {Q}_\alpha^{\text{RP}}$, involves the exchange
    of energy between different reservoirs, while
    resonant heating (RH), associated with
     $\dot {Q}_\alpha^{\text{RH}}$, involves the exchange
     of energy between different modes of the same
     environment. The total number of excitations
     in the reservoirs is conserved due to these two
     mechanisms.}
    \label{fig:res_proc}
\end{figure}

Using the explicit form of the Hamiltonians,
it is relatively simple to express the heat currents in
the stationary regime in terms of the Fourier
components of the Green function.  Again,
we do not present the detailed calculation but
simply describe the ingredients needed to obtain it
(the full derivation can be found in Ref. \cite{freitas2017}).
The heat current in the stationary regime can be
fully expressed in terms of the covariance matrix
of $\cals$. Of course, this comes as no surprise
since the state of $\cals$ is Gaussian in the
stationary limit (and therefore
it is fully determined by its
covariance matrix). The average heat
flow has a particularly simple form:
\beq
\dot {Q}_\alpha=\overline{V(t)\sigma_{xp}(t)},\nonumber
\eeq
where $\sigma_{xp}(t)$ is the position-momentum
correlation function of $\cals$, defined
as $\sigma_{xp}(t)=\langle{\{x(t),p(t)\}\rangle}/2$
(the overbar in the previous equation
indicates the average of the product between
$V(t)$ and $\sigma_{xp}(t)$
over a period of the oscillation).
In the stationary regime, when
the memory of the initial state is lost,
the correlation function can
be fully expressed in terms of
the Green function $G(t,t')$
as follows:
\beq
\sigma_{xp}(t)=\frac{1}{2}\int_0^\infty\int_0^\infty
dt_1 dt_2 G(t,t_1)\nu(t_1,t_2)\dot G(t,t_2),\nonumber
\eeq
where the noise kernel of the environments is
defined as
\beq
\nu(t)=\int_0^\infty d\omega \sum_\alpha I_\alpha(\omega)
\coth(\omega/2k_BT_\alpha)\cos(\omega t).\nonumber
\eeq
Replacing the previous expressions for $G(t,t')$,
for $V(t)$ and using the set of equations
satisfied by the Floquet
coefficients $A_k(\omega)$ we can derive a
particularly simple and
physically appealing formula for $Q_\alpha$.
In fact, in Ref \cite{freitas2017} it
was shown that the heat current can
always be expressed as the sum of three
terms
\begin{equation}
\dot Q_\alpha=\dot Q_\alpha^{RP}+\dot Q_\alpha^{RH}+
\dot Q_\alpha^{NRH}.\nonumber
\end{equation}
The explicit form of the
resonant pumping (RP), the
resonant heating (RH) and the non--resonant
heating (NRH) contributions to the average heat
fluxes will be given below and will enable
us to understand the various thermodynamical
processes involved. Thus, the first term is:
\begin{equation}
\begin{split}
    \dot{Q}_\alpha^{\text{RP}} &=
    \sum_k \int_{0'}^\infty d\omega \;
    \left( \omega \;
    p^{(k)}_{\beta,\alpha}(\omega) \; N_\alpha(\omega)\right.\\
    & \left. - (\omega+k \omega_d) \; p^{(k)}_{\alpha,\beta}(\omega) \; N_\beta(\omega)
    \right)
    \label{meq:heat_rp}
\end{split}
\end{equation}
where
$N_\alpha(\omega) = (e^{\omega/T_\alpha}-1)^{-1}$
is the Planck distribution and
\beq
p_{\alpha,\beta}^{(k)}(\omega)=\frac{\pi}{2}
I_\alpha(|\omega+k
\omega_d|) I_\beta(\omega) |A_k(\omega)|^2
\eeq
is a positive number,
proportional to the
probability for the system to couple a
mode $\omega$ in
$\cale_\beta$ with mode
 $|\omega+k\omega_d|$ in
$\cale_\alpha$. The diagram in
Figure \ref{fig:res_proc} describe
the processes involved in
$\dot Q_\alpha^{RP}$:
excitations are transported between reservoirs
after the resonant absorption (or emission)
of an energy $k\, \omega_d$, a multiple of
the quantum of energy of the
driving. The origin of the resonant heating
$\dot Q_\alpha^{RH}$ is similar but in this case,
as shown in the same Figure,
the excitations are transported within the
reservoir $\cale_\alpha$, which always
results in
heating. The corresponding formula is simply
\begin{equation}
    \dot{Q}_\alpha^{\text{RH}} =
     - \sum_{k>0} \int_{0'}^\infty
    d\omega \; k \omega_d  \;
    p^{(k)}_{\alpha,\alpha}(\omega) \;
    (N_\alpha(\omega)-N_\alpha(\omega+k\omega_d)).
    \label{meq:heat_rh}
\end{equation}
Two points are worth noticing: a)
the lower limit of
integration in the expression RH  current
is not $\omega = 0$ but
$\omega =0'=\max\{0,-k \omega_d \}$.
This is because when $k<0$,
the arrival mode exists only if
$\omega\ge -k\omega_d$. For negative
values of the integer $k$,
the low frequency part of the spectrum
plays no role.
As we will see, the low frequency part of the
spectrum is
crucial in the non resonant processes.
b) The
negativity of
$\dot{\bar Q}_\alpha^{\text{RH}}$
(which is the reason why this term is
always associated with heating)
relies on the fact that the Planck distribution
decreases with frequency. In fact, population
inversion can change the sign of this term.

\begin{figure}[ht]
    \centering
    \includegraphics[width=.35\textwidth]{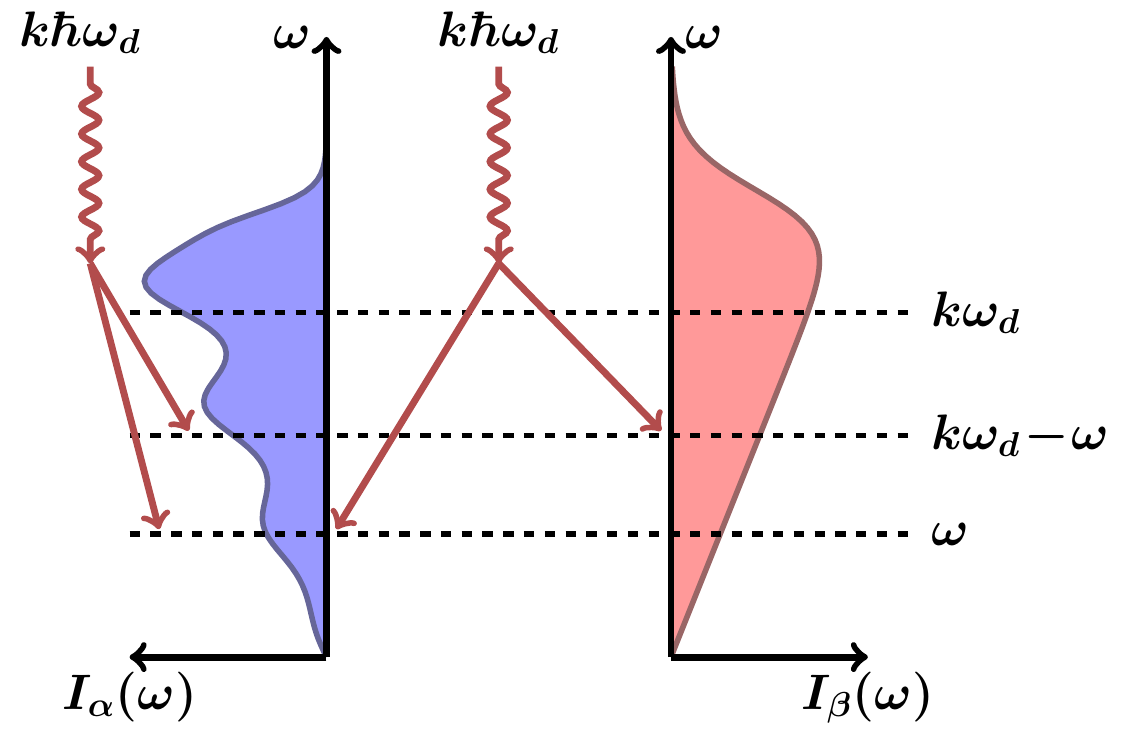}
    \caption{Non Resonant processes affecting the heat
    flow in and out of each reservoir. They can involve
    either modes of two different reservoirs or
    modes of the same reservoir. In all cases,
    excitations are
    created in pairs.}
    \label{fig:nonres_proc}
\end{figure}

The last term in the heat flow fully
contains the contribution of low frequencies.
It is  the non resonant heating term
$\dot {Q}_\alpha^{\text{NRH}}$ which reads:
\begin{equation}
\begin{split}
    & \dot{Q}_\alpha^{\text{NRH}} =
    - \sum_{k>0} \int_{0}^{k\omega_d}
     d\omega \; k \omega_d \; p_{\alpha,\alpha}^{(-k)}(\omega) \;
     \left(N_\alpha(\omega)\!+\!\frac{1}{2}\right) -\\
     & - \sum_{k>0} \int_{0}^{k\omega_d}
     d\omega \; (k\omega_d\!-\!\omega)\;p_{\alpha,\beta}^{(-k)}(\omega) \;
     \left(N_\beta(\omega)\!+\!\frac{1}{2}\right)-\\
     & - \sum_{k>0} \int_{0}^{k\omega_d}
     d\omega \; \omega \; p_{\beta,\alpha}^{(-k)}(\omega) \;
     \left(N_\alpha(\omega)\!+\!\frac{1}{2}\right)
        \label{meq:heat_nrh}
\end{split}
\end{equation}
The non resonant heating does not involve the transport
of excitation but the creation of excitation pairs. These processes
are depicted in Figure \ref{fig:nonres_proc}. In that
Figure we see that these
pairs can be
created either in different environments or in the same
environment giving rise to the three contributions shown in the
above formula. The interpretation of the equation is
transparent: For each Floquet term (i.e., for
each integer $k$), when an excitation with
frequency $\omega$ is emitted into one of the reservoirs
the second excitation must have a
frequency equal to $k\omega_d-\omega$.
The probability
for creating the first excitation has a stimulated
contribution (proportional to the mean number of excitations
already present in that mode) and also a spontaneous
contribution. For each of the possible emission
channels, the energy deposited in $\cale_\alpha$ is
either $\omega$, $k\omega_d-\omega$ or
$k\omega_d$. It is worth
stressing the fact that Eq. (\ref{meq:heat_nrh}) is
an exact formula, which is
valid both for weak and strong coupling
\cite{freitas2017}.

\subsection{The fundamental limit for cooling}
Clearly, $\dot Q_\alpha^{NRH}\le 0$ and therefore it corresponds to
the heating of all reservoirs. Moreover, when the temperatures
of all reservoirs vanish, this is the only term contributing
to the heat flow. As a consequence,
this term always becomes dominant at
sufficiently low temperatures. For this reason,
it is responsible for the
validity of the third law of thermodynamics.
Thus, as discussed in \cite{freitas2017},
if the non resonant heating term
is not taken into account
(which is clearly erroneous at low temperatures)
the third law could indeed be
violated, as suggested recently\cite{kolavr2012}.

Identifying the process of pair creation as the one
imposing the fundamental limit for cooling may sound a
bit exotic. As far as we know, this process is never
mentioned in the most relevant texts on
laser cooling \cite{cirac1992,eschner2003}
as the one to blame for the fact that
such type of cooling process cannot
reach zero temperature.
On the other hand,
pair production through parametric driving
is a well known phenomenon that
has been identified as relevant in many other contexts which
seem to be unrelated to the one we are analyzing. For example,
the spacetime dynamics induces cosmological pair
creation \cite{birrell1984},
which play an important role
in the history of our Universe. In fact, pair creation due to
parametric driving is particularly relevant during the
relating stage of  inflationary models of the early
Universe \cite{mazzitelli1989}).
Also, the motion of
the mirrors that form a cavity excites the vacuum
and induces the
creation of pairs of photons inside the cavity. This
process, that is known as the dynamical Casimir effect
(DCE) \cite{dodonov2010,birrell1980,birrell1984} was recently
studied both with theoretical and experimental
methods \cite{dodonov2010,wilson2011,lahteenmaki2013}. The analogy between our
analysis and the typical setup used to analyze the
DCE is clear: In fact, when there is a single reservoir,
our equation (\ref{meq:heat_nrh}) is identical
to the one describing the power spectrum of created pairs in the DCE.
We can interpret this as a consequence of the fact that, in
our case, the driving of $\cals$ effectively act as some
 sort of moving boundary conditions for the environmental
 modes and, in this way, it induces pair creation.
This is a rich analogy and we will exploit it
below, where we will show
that pair creation is the natural process to be blamed
for imposing the lowest achievable temperature
in the cooling of a single motional mode as it is
typically done in laser cooling.

\section{Cooling a single motional mode}
The previous expressions are valid for
arbitrary spectral densities. We will apply them
here to analyze the limits for cooling of a single
motional mode, whose frequency we denote
as $\omega_m$. For this, we consider the spectral
density of $\cale_A$ to be such that
$I_A(\omega)=\tilde I_A\, \delta(\omega-\omega_m)
$, where $\tilde
I_A$ is a constant. In this case,
the relevant processes for cooling and heating
of $\cale_A$ are displayed in Figure \ref{fig:laser_cooling_proc}
(where a spectral density peaked about $\omega_m$ was
used for illustrative purposes). Clearly, the RH
process is absent since $\cale_A$ consists only of a
single mode. The lowest achievable temperature
can be estimated by analyzing the condition under
which the heating and cooling terms balance
each other. Using the above equations it is
simple to compute this ratio as
\begin{equation}
\left| \frac{ \dot{Q}^\text{RP}_A}{\dot Q_A^{NRH}} \right|
   = \frac{\bar n}{1+\bar n}
   \frac{\sum_{k>0}I_\beta(k\omega_d+\omega_m) |A_k(\omega_m)|^2}{\sum_{k>0} I_\beta(k\omega_d-\omega_m)
|A_{-k}(\omega_m)|^2},
    \label{eq:ratiofull}
\end{equation}
where $\bar n=N_A(\omega_m)$ is the average number
of excitations in the motional mode (the one that
is being cooled). In order to simplify our analysis,
 we neglected the heating term appearing in the resonant
 pumping current $\dot Q^{RP}$. By doing this, we study
 the most favorable condition for
 cooling, assuming that the pumping of
excitations from $\cale_B$ into
$\cale_A$ is negligible. This is equivalent
 to assuming that the temperature of $\cale_B$ is
 $T_B\simeq 0$. Although this is a reasonable approximation
 in many cases (such as the cooling of a single trapped
 ion) we should have in mind that by doing this, the
 limiting temperature we will obtain should be actually
 viewed as a lower bound to the actual one.
 Thus, the condition defining the lowest bound
 is that the ratio between
the RP and NRH currents is
of order unity. Using the previous expressions,
it is simple to show that this implies that
\begin{equation}
 \frac{1}{\bar n}=
 \frac{\sum_{k>0}I_B(k\omega_d+\omega_m) |A_k(\omega_m)|^2}
{\sum_{k>0} I_B(k\omega_d-\omega_m) |A_{-k}(\omega_m)|^2}-1.
     \label{eq:nfull}
\end{equation}
To pursue our analysis, we need an expression
for the Floquet coefficients $A_k(\omega)$.
This can be obtained under some simplifying
assumptions. In fact,
if the driving is harmonic
(i.e. if $V(t)=V_0 + V(e^{i\omega_dt}+e^{-i\omega_dt})$)
and that its amplitude is small (i.e. if
$V\ll V_0$), we can
use perturbation theory to compute the
Floquet coefficients to leading order in $V$.
In fact, by doing this we find that the only
non vanishing terms are the ones corresponding
to $k=\pm 1$ which read
\beq
A_{\pm 1}(\omega_m)\approx -
g(i(\omega_m\pm\omega_d))V\,
g(i\omega_m).
\eeq
Using this, we find that
\begin{equation}
 \frac{1}{\bar n} =
   \frac{I_B(\omega_d+\omega_m)|g(i(\omega_d+\omega_m))|^2}
{I_B(\omega_d-\omega_m)
|g(i(\omega_d-\omega_m))|^2}-1
\label{eq:nmink1}
\end{equation}
The final step of our derivation requires
the use of an expression for $g(s)$ (the
propagator of the undriven oscillator).  For this
we use a semi phenomenological approach
by simply assuming that, in the absence
of driving, the coupling with
the reservoirs induces an exponential
decay of the oscillations of $\cals$.
In this case, we can simply write
$g(i\omega)=1/((\omega-i\gamma/2)^2-\omega_0^2)$,
where $\gamma$ is the decay rate and $\omega_0$ is
the renormalized frequency of $\cals$.
As mentioned above,
this same expression is obtained if we assume
that $\cals$ behaves as if it were
coupled with a single ohmic environment
(this is indeed a reasonable
assumption in many cases but it certainly
requires that the back action of
$\cale_A$ on $\cals_S$ is negligible in the
long time limit). Then, we can simply write
that
\beq
|g(i\omega)|^2 =\frac{1}{(\omega^2 -\omega_0^2+\gamma^2)^2
+4\gamma^2\omega_0^2}.
\eeq
Using this, we find the following
expression for the lowest achievable number
of excitations in $\cale_A$:
 \begin{equation}
  \frac{1}{\bar n} =
   \frac{I_B(\omega_d+\omega_m)}
   {I_B(\omega_d-\omega_m)}
   \frac{((\omega_d-\omega_m)^2-\omega_0^2
+\gamma^2)^2+4\omega_0^2\gamma^2}
{((\omega_d+\omega_m)^2-\omega_0^2
+\gamma^2)^2+4\omega_0^2\gamma^2}-1.
     \label{eq:nmingeneral}
\end{equation}
This relatively simple formula, is an important
result of our paper. It show that the
asymptotic value of
$\bar n$ depends on four parameters:
$\omega_0$,
$\omega_m$, $\gamma$ and $\omega_d$.
The first three of them characterize the system while
$\omega_d$ can be adjusted to
minimize $\bar n$. As shown below, the optimal
driving frequency depends on the other parameters in an interesting
manner. In particular, we will study two physically
important regimes where the decay rate
satisfies either that $\gamma\ll\omega_m$ (the
limit of resolved sidebands) or
$\gamma\gg\omega_m$
(the limit of non resolved sidebands). In both
cases we will show that our expression for
the limiting temperature coincides with the
one obtained when studying the limit for laser
cooling.

\subsection{The limit of sideband resolved
laser cooling}
Let us analyze the conditions under which it is
possible to achieve ultra low
temperatures, i.e. temperatures such that
$\bar n\ll 1$. In that case, it is simple to show
that the optimal driving frequency is the
one for which the denominator appearing
in the expression for $1/\bar n$ is minimal.
Thus, assuming that
the ratio $I_B(\omega_d-\omega_m)/I_B(\omega_d+
\omega_m)$ is a slowly varying function of
$\omega_d$ (which is certainly the case when
$\omega_d\gg\omega_m$ such as in the case
of a trapped ion), the optimal driving frequency
is such that
$\omega_d=\sqrt{\omega_0^2-\gamma^2}
-\omega_m$. In this case, the lowest achievable
temperature is such that
\begin{equation}
\begin{split}
\bar n&=
\frac{\gamma^2\omega_0^2}{4\omega_m^2\omega_d^2+\omega_0^2\gamma^2}
\frac{I_\beta(\omega_d-\omega_m)}
   {I_\beta(\omega_d+\omega_m)}\\
&\approx
\frac{\gamma^2}{4\omega_m^2}
\frac{\omega_0^2}{(\omega_0-\omega_m)^2}
\frac{I_B(\omega_0-2\omega_m)}
   {I_B(\omega_0)},
\end{split}
\end{equation}
where the approximation $\gamma\ll\omega_0$
was used to obtain the last expression. In
this case $\omega_d\approx\omega_0-
\omega_m$, which implies that the
optimal situation is achieved
when the driving is resonant with the red
detuned lateral sideband of the carrier frequency
$\omega_0$.
Clearly, the condition $\bar n\ll 1$ can only
be satisfied
if $\gamma\ll\omega_m$, which is nothing
but the condition
for resolved sidebands. Notably, the above
expression for $\bar n$
coincides with previously known formulae
for the lowest achievable temperature in
sideband resolved laser cooling, which were
obtained by completely different methods.
In fact, in that case a further simplification
is possible as laser cooling is typically
done in the optical regime for $\omega_0$
and $\omega_d$ (which are therefore much
larger than $\omega_m$, which is typically
in the r.f. domain). Then,
the ratio $I_B(\omega_d-\omega_m)/I_B(\omega_d
+\omega_m)$
is of order unity and, therefore, we have that
the lowest temperature is such that
$\bar n=\gamma^2/4\omega_m^2$. Our
formula is clearly more general and can be
applied in other cases, as discussed below.

\subsection{The limit of Doppler cooling}
Let us now study a different regime, where
the cooling does not achieve ultra low
temperatures (i.e., the condition $\bar n\ll 1$ is
not satisfied).
If we neglect the contribution arising from the
spectral densities (which, as above, is a reasonable
approximation whenever $\omega_0\gg\omega_m$, like in the optical
regime) the asymptotic value of the occupation
number is
\begin{equation}
\bar n=\frac{
\left((\omega_d+\omega_m-\omega_0)^2+\gamma^2\right)
\left((\omega_d+\omega_m+\omega_0)^2+\gamma^2\right)
}{8\omega_d\omega_m
\left(\omega_0^2-\omega_m^2-\gamma^2-\omega_d^2
\right)}\nonumber
\end{equation}
Cooling is possible only if the condition
$\omega_0^2>\omega_m^2+\gamma^2+\omega_d^2$
is satisfied (notice that this implies that $\omega_0$ must be larger than $\omega_d$,
$\omega_m$ and $\gamma$).
The above expression (which is valid
whenever the spectral density varies slowly with $\omega_m$) can be used to study the case
where $\gamma\gg\omega_m$, which physically
very important. In fact, this regime corresponds
to the Doppler cooling limit, where the sidebands
are not resolved. In this case, the optimal
driving frequency can be shown to be
$\omega_d=\omega_0-\gamma$ and the
limiting temperature turns out to be
\begin{equation}
\bar n=\frac{\gamma}{2\omega_m}\nonumber
\end{equation}
In this case, the cooling condition is satisfied
whenever $2\gamma\omega_d\ge\omega_m^2$
(analogously, in the sideband
resolved case the cooling condition requires
that $2\omega_m\omega_d\ge\gamma^2$, which
is naturally satisfied). Thus, by applying a single
formula in two different situations we obtained the
limiting temperatures in the most relevant regimes
for laser cooling. This seems to indicate that
the mechanism that is responsible for stopping
laser cooling is pair creation. Let us analyze
this in some detail.

\subsection{The role of pair creation in
laser cooling}

\begin{figure}[ht]
    \centering
    \includegraphics[width=.35\textwidth]{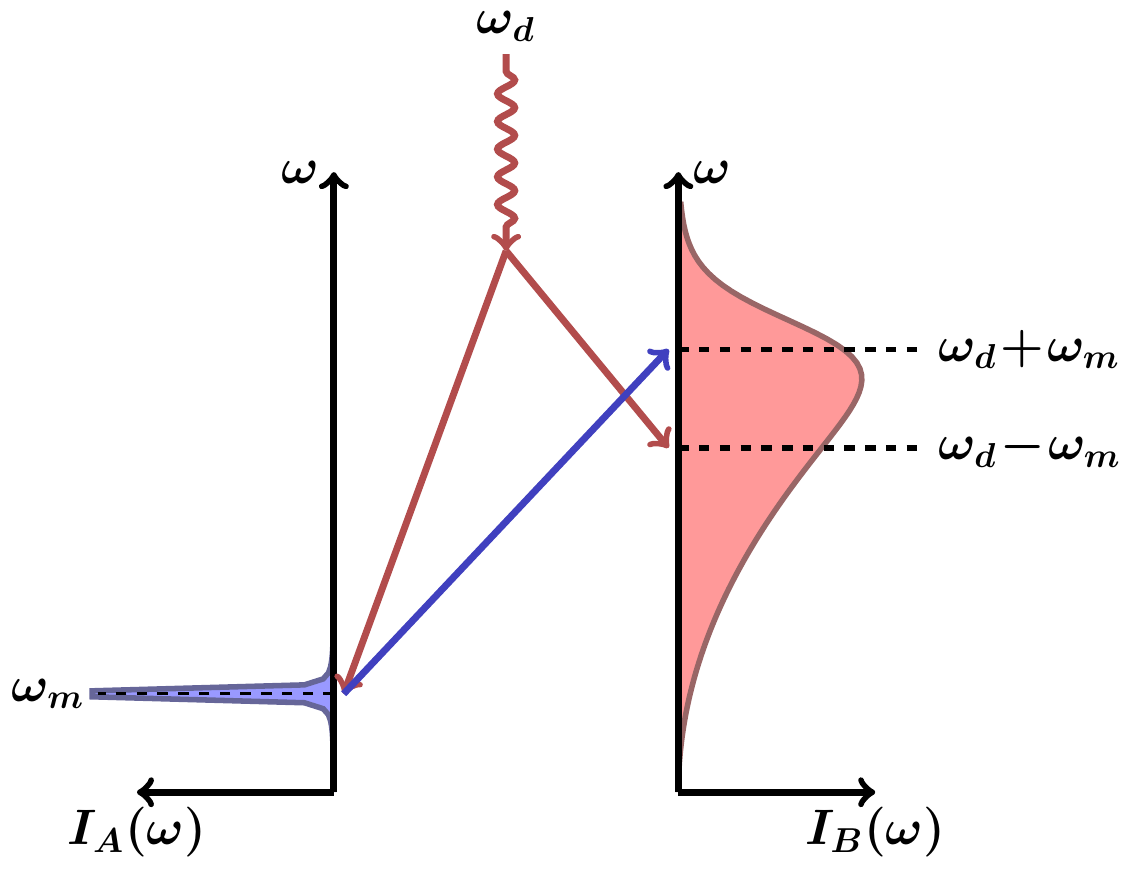}
    \caption{Relevant processes contributing to the heat current of reservoir
    $\cale_A$ when $\omega_d>\omega_m$ and $T_B\simeq0$.}
    \label{fig:laser_cooling_proc}
\end{figure}

\begin{figure}[ht]
    \centering
    \includegraphics[width=.45\textwidth]{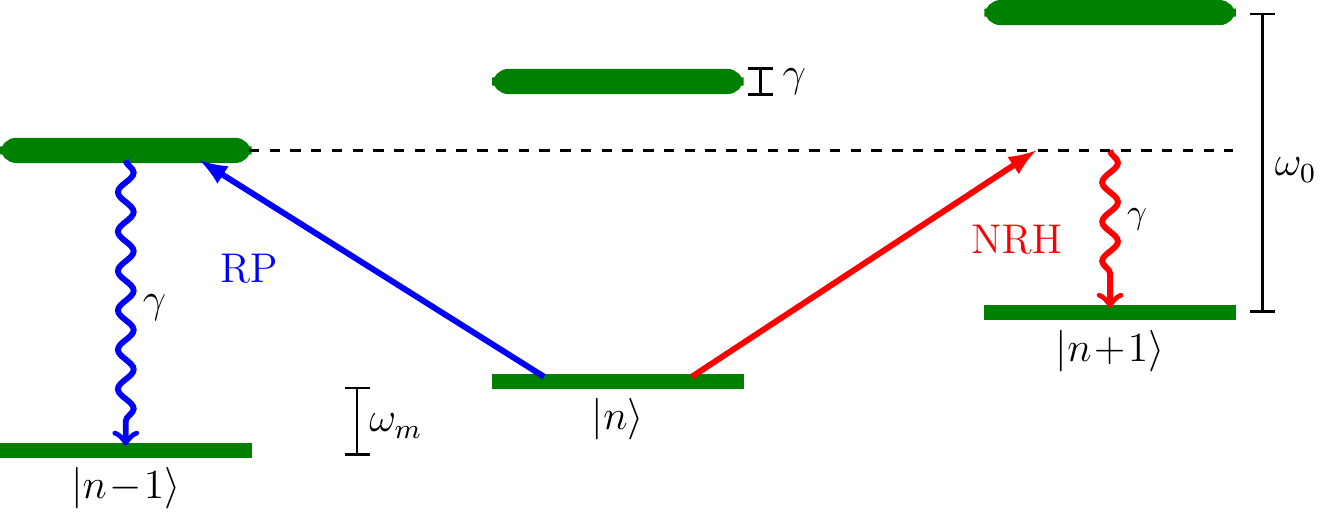}
    \caption{Usual depiction of the staircase of energy
    levels and the transitions between them
    involved in
    Doppler and sideband laser cooling (actually, there are other non resonant processes in
play, see \cite{eschner2003})}
    \label{fig:sideband_ladder}
\end{figure}

According to our analysis,
the origin of the lowest achievable temperature
in the class of refrigerators we analyzed is imposed
by pair creation from the driving.  This is certainly
not the typical explanation for the reason why
laser cooling stops. However, we will see now
that pair creation has a natural role in laser cooling.
Thus, we will see that
pair creation is certainly not an exotic but an
essential process in this case. The
relevant processes that play a role in the
resonant pumping and non resonant heating
currents are
shown in Fig. \ref{fig:laser_cooling_proc}
(for $\omega_d>\omega_m$).
Thus, the resonant pumping of energy
out of $\cale_A$ (blue arrow in Figure \ref{fig:laser_cooling_proc}) corresponds to
a removal of a motional excitation (a phonon)
and its
transfer to the photonic environment. A phonon
with frequency $\omega_m$ disappears
in $\cale_A$ and a photon with frequency $\omega_0$
appears in $\cale_B$. This is possible by absorbing
energy $\omega_d=\omega_0-\omega_m$
from the driving. This process is usually
visualized in a different way in the standard
literature of laser cooling \cite{eschner2003},
as shown in Fig. \ref{fig:sideband_ladder}.
This Figure shows the energy levels of the
combined system formed by $\cale_A$ and $\cals$.
In our case, both systems are oscillators and
each one of them has an infinite number of
energy levels. However,
we only pay attention to the lowest levels
of $\cals$. Thus, the resonant pumping process
(RP) takes the system from the lowest energy
level of $\cals$ with $n$ phonons into the
excited level of $\cals$ with $(n-1)$ phonons.
Then, as $\cals$ is coupled to the
environment $\cale_B$, it decays from the excited $\ket{e}$
to the ground ($\ket{g}$) state by emitting an
excitation (a photon) in $\cale_B$,
whose frequency
is $\omega_0$. This is the key process
responsible for sideband resolved laser
cooling. The system is cooled because resonant
pumping forces the combined $\cale_A-\cals$
system to move down in the staircase of energy
levels.

However, if resonant pumping were the only
relevant process, the above argument would
induce us to conclude that laser cooling could
achieve zero temperature: by going down the
staircase of energy levels, $\cals$ would end
up in its ground state and the motional
mode would end up with $n=0$ phonons.
The reason why this does not happen is the
existence of non resonant
heating. This process is described as NRH in Fig. \ref{fig:sideband_ladder}.
It corresponds to the creation of a pair
of excitations consisting of a
phonon and a photon. The phonon, has frequency
$\omega_m$ while the photon should have frequency
$\omega_d-\omega_m$. We may choose to
describe this pair creation process as a sequence of
heating transitions that moves the combined $\cals$--$\cale_A$
system up along the staircase of energy levels. This can
be done as follows: Suppose that we start
from $n$ phonons in the motional state and
$\cals$ in the ground state $\ket{g}$. Then, $\cals$
can
absorb energy $\omega_d$ from the driving and
jump into a virtual state from which it can decay
back into $\ket{g}$ but with a motional
state with $n+1$ phonons.
This heating transition has the net effect
of creating a phonon and  emitting a photon.
As we wrote before, and we stress again,
laser cooling stops (in this sideband resolved limit)
when the resonant cooling transitions are
compensated by non resonant heating transitions
where energy is absorbed from
the driving and  is split between two excitations:
one in the motional mode (a phonon) and one
in the environmental mode (a photon). As a
consequence of the non resonant transitions,
the motion heats up. The
limiting temperature is achieved when the
resonant (RP) and non resonant (NRH) processes
balance each other.

Of course, the way in which we are describing
the processes involved in laser cooling (both
the cooling and the heating transitions) is
not the standard one but provides a new
perspective and some new
predictions, which we will discuss below.

\subsection{Ultra low temperatures for
structured reservoirs}

A first generalization of the standard results
for the limiting temperatures achieved in
laser cooling arises if we can manipulate
the environmental spectral density. In that
case, new results emerge, which we analyze
now. In fact, the properties of
spectral density of the
reservoir may be very relevant in establishing
the lowest achievable temperature in some
cases. Thus, this is because of the
presence of the ratio  $I_B(\omega_d-\omega_m)/I_B(\omega_d+\omega_m)$
in Eq.
(\ref{eq:nmingeneral}). As we will see,
this ratio
may play an interesting role in certain cases. If the spectral density is a monotonic function
of the frequency (at least in a relevant band
around $\omega_d$),
the ratio is always smaller than unity.  As we
mentioned above, when $\omega_m$ is much
smaller than $\omega_d$, the ratio is close
to unity and can be neglected. However,
when $\omega_m$ is of the same order of
magnitude than $\omega_0$, the ratio
may be very small and
substantially modify the minimal occupation
number that the cooling mechanism can achieve.

Let us analyze this now. We take
$\omega_0$ and $\omega_m$ to be
similar. As before, we assume we are in an underdamped regime where
$\omega_0\gg\gamma$. Therefore, we
also have $\omega_m\gg\gamma$ and the
resolved sideband condition is satisfied.
Thus, the
 optimal driving frequency
is $\omega_d = \omega_0 - \omega_m$
and the minimum occupation number is
given above by Eq. (\ref{eq:nmingeneral}).
If the spectral density is such
that $I_B(\omega) \propto \omega^\kappa$, then Eq.
(\ref{eq:nmingeneral}) becomes:
\begin{equation}
    \bar n_\text{min} = \frac{1}{4}\left(\frac{\Gamma}{\omega_m}\right)^2
    \frac{(1-2\omega_m/\omega_0)^\kappa}{(1 - \omega_m/\omega_0)^2}.
\end{equation}
For $\kappa \geq 1$ (Ohmic and super-Ohmic spectral densities),
the factor $f = (1-2\omega_m/\omega_0)^\kappa/(1 - \omega_m/\omega_0)^2$ is always
less than unity. As a consequence,
the minimum achievable temperature is lower than the
one given by the standard formula
for sideband cooling. Instead, for
sub-Ohmic spectral densities,
the condition $f<1$ is satisfied
only when
$\omega_m/\omega_0$ is larger than a
critical value. For example, if
$\kappa=1/2$, we have $0<f<1$ only if $0.457<\omega_m/\omega_0<0.5$. In turn,
for highly sub-Ohmic environments
with $\kappa\ll 1$, the condition $f<1$ is
satisfied only in a very narrow band (defined
as $1/2(1-1/2^{2/\kappa})<
\omega_m/\omega_0<1/2$.

If the environmental spectral density is such
that $I(\omega=0)=0$, then
the lowest value of $\bar n$ obtained
from Eq. \ref{eq:nmingeneral}
tends to zero when
$\omega_m \to \omega_0/2$. In fact,
in this case the
pair creation mechanism described above
is suppressed (because the excitation would
have to be created at very low frequencies,
where the environment has no available modes).
This result may lead us to the erroneous
conclusion than zero temperature could
be achieved. Indeed, this is not the case
because when $\omega_d=\omega_m$,
the non resonant heating current is dominated
by the next to leading order term in the Floquet
index (i.e., $k=\pm 2$). Thus, in this case the
pairs of excitations are created by absorbing
energy $2\omega_d$
from the driving. In fact, this energy is
enough to create an excitation in the motion
and another excitation in the environment at
frequency $\omega_m$.
Of course, this process is of higher order
in the driving amplitude $V$. In this case, going back
to Eq. \ref{eq:nfull} we can estimate the
correct limiting temperature. Thus, using
the fact that, when $\omega_d=\omega_m$
we have $A_{-2}(\omega_m)=g(-i\omega_m)
V A_{-1}(\omega_m)$ we obtain
\begin{equation}
\bar n|_{\omega_0=2\omega_m}
\approx
\frac{\gamma^2}{\omega_m^2}
\frac{I_B(\omega_m)}
   {I_B(2\omega_m)}
\frac{64 V^2}{9\omega_m^4}.
\end{equation}
Therefore, Eq. (\ref{eq:nmingeneral}) is valid only if
$\omega_m$ is not too close to $\omega_0/2$. Otherwise, higher order processes
must be taken into account.

These considerations are not relevant
for the majority of sideband cooling implementations, where the motional
frequency $\omega_m$ is several orders of magnitude lower than the
central system frequency
$\omega_0$\cite{leibfried2001,vuletic1998,teufel2011}. However,
they might be of interest for systems of superconducting qubits
coupled with radio frequency cavities. In that case,
one could use the qubit to pump energy away
from one cavity dumping it in a second one.
In this type of systems the
frequencies of these two objects can be selected by design, and they can be of
the same order (in fact, in this
type of system, the DCE was already observed\cite{wilson2011,lahteenmaki2013}).

\section{Power spectrum of the emitted radiation: }
As was explained in the previous sections,
in the usual presentation of laser cooling the heating
mechanisms preventing the perfect
preparation of the motional ground state are
understood as inelastic scattering events
involving transitions to virtual
electronic levels followed by spontaneous emissions\cite{eschner2003}. When one of these
processes takes place, the overall
effect is the creation of a motional
excitation and the emission of a photon
with the frequency reduced by
$\omega_m$ with respect to the
incident radiation.
\begin{figure}[ht]
    \centering
    \includegraphics[width=.4\textwidth]{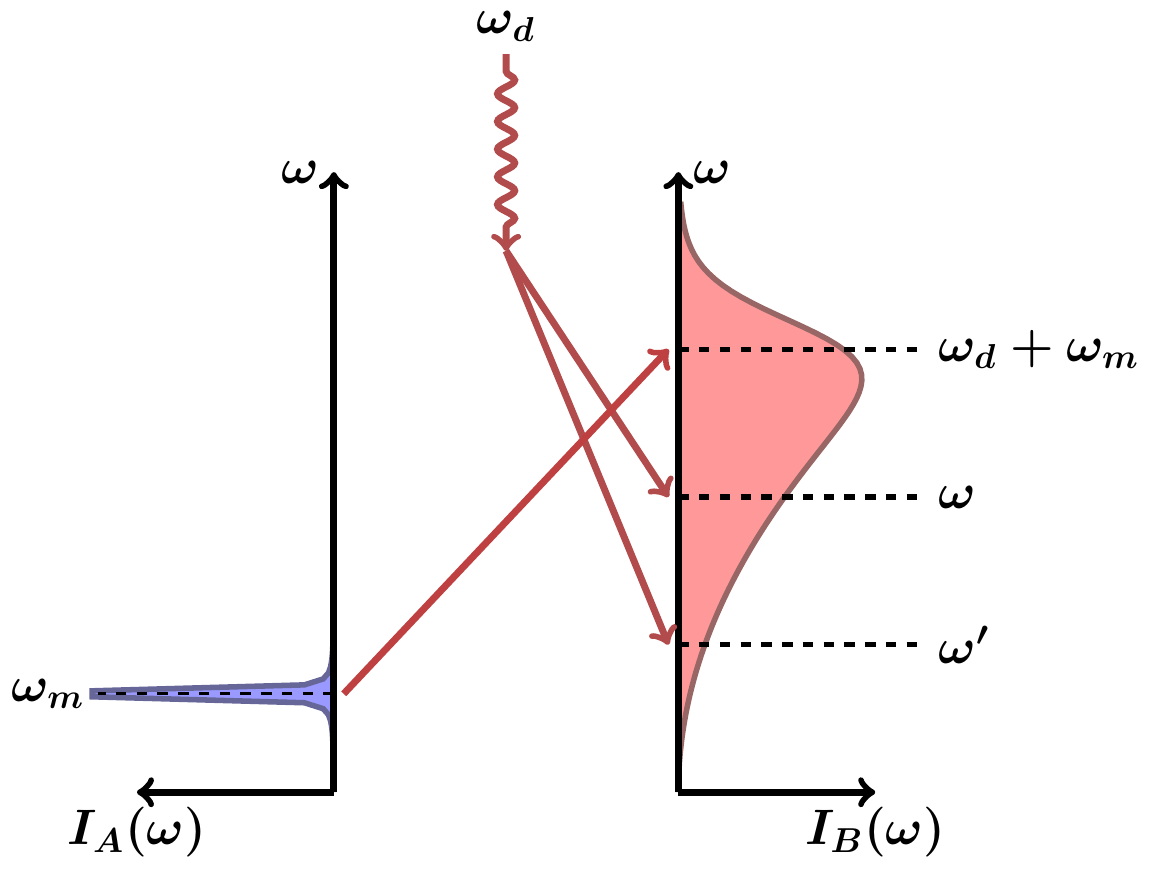}
    \caption{Relevant processes contributing to the heat current of reservoir
    $\cale_B$ when $\omega_d>\omega_m$ and $T_B\simeq0$. Pairs of photons are created
    at frequencies $\omega$ and $\omega'$ such that $\omega + \omega' = \omega_d$.}
    \label{fig:cooling_EM_heat}
\end{figure}
 From our point of view,
this process can also be
understood as a particular case of the pairs creation
mechanism, analogous to the one present in the
DCE.
Thus, the motional excitation and
the emitted photon, whose frequency is
$\omega_d -
\omega_m$ are analogous to a dynamical
Casimir pair. However, there is another
aspect of the cooling process in which the
pair creation process plays a role. This can be
seen by analyzing the heat current entering the
reservoir $\cale_B$, which
represent the electromagnetic field. In fact, during the
cooling process there will be three types of photons
which will be emitted in $\cale_B$. Firstly, we will find
photons with the carrier frequency $\omega_0$ which
are produced during the resonant cooling transitions
(where a phonon is transformed into a photon by
absorbing energy from the driving). Secondly, we will
find the photons emitted during the heating transitions which, as
described above, have frequency $\omega_0-2\omega_m$.
These two processes were described above and
are associated with the diagrams presented in
Figure \ref{fig:laser_cooling_proc} and \ref{fig:sideband_ladder}.
But there will be a third class of photons which are
emitted in pairs directly from the driving.

\begin{figure}[ht]
    \centering
    \includegraphics[width=.4\textwidth]{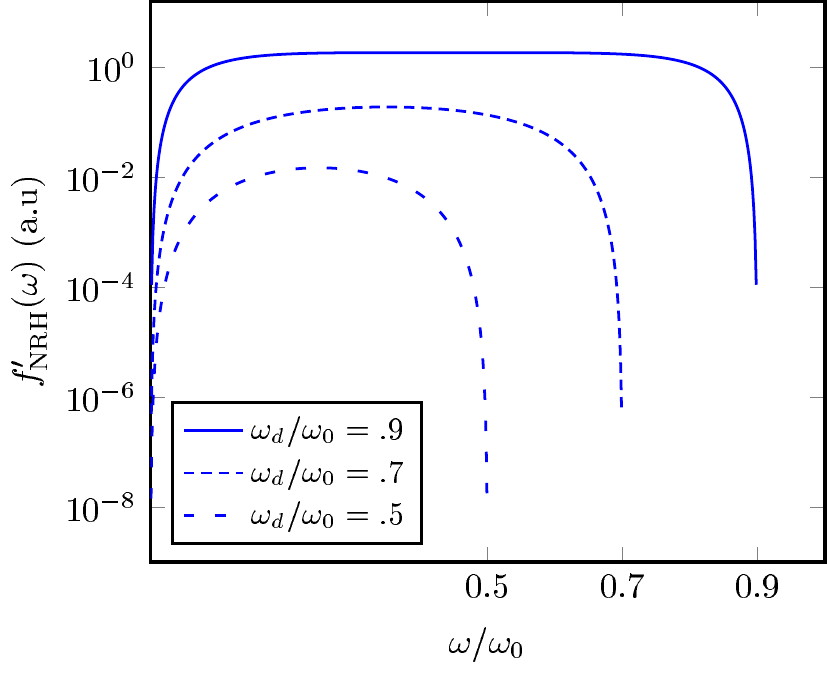}
    \caption{Power spectrum for the creation of photon pairs directly into the
    electromagnetic field (for $I_B(\omega) \propto \omega^3$).}
    \label{fig:power_spectrum}
\end{figure}
\begin{figure}[ht]
    \centering
    \includegraphics[width=.4\textwidth]{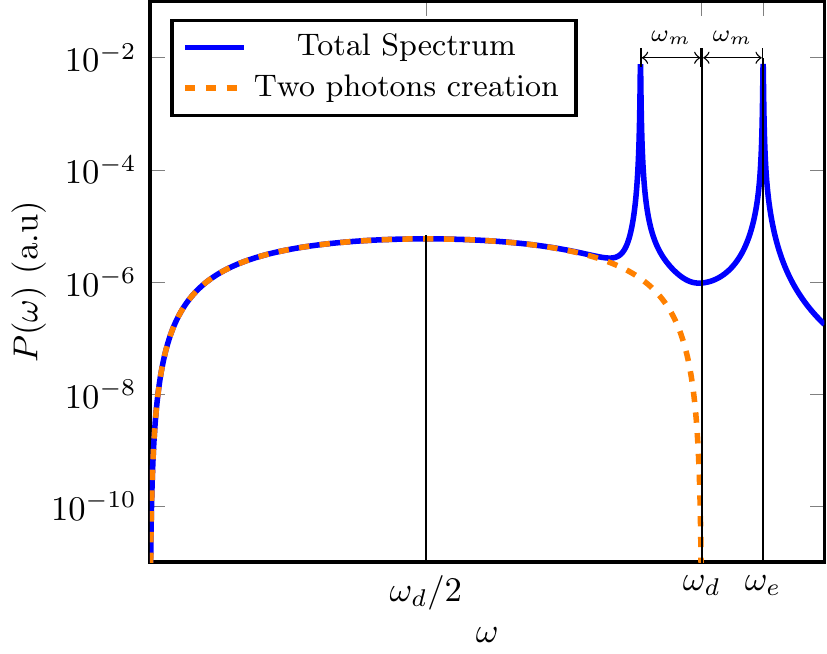}
    \caption{ Full power spectrum for the heat current entering the
        electromagnetic field (for $I_B(\omega) \propto \omega^3$). }
    \label{fig:power_spectrum_full}
\end{figure}

All the above mentioned contributions to the
electromagnetic radiation during the cooling process can
be analyzed from our previous formulae. In fact,
the power spectrum of the energy dumped
into the EM field can be read from
the integrand of Eqs. (\ref{meq:heat_rp}), (\ref{meq:heat_rh}) and
(\ref{meq:heat_nrh}). They precisely contain the three
contributions we mentioned above. The first one comes from the
resonant pumping (RP) of energy from
reservoir $\cale_A$ to $\cale_B$, which creates photons at
frequency $\omega_d + \omega_m$. Thus, if $f_\text{RP}(\omega)$ is the number of
photons per unit of time created at frequency $\omega$ by this process, we
have, for $\omega>\omega_d$:
\begin{equation}
    f_\text{RP}(\omega) = \frac{\pi}{2} I_B(\omega) I_A(\omega-\omega_d)
    |A_1(\omega-\omega_d)|^2
    N_A(\omega-\omega_d)
    \label{eq:f_RP}
\end{equation}
In the same way, the number of photons per unit of time created at frequency
$\omega$ by the pair creation of Figure \ref{fig:laser_cooling_proc} is:
\begin{equation}
    f_\text{NRH}(\omega) = \frac{\pi}{2}
    I_B(\omega) I_A(\omega_d-\omega) |A_{-1}(\omega)|^2
    \left[N_A(\omega_m)+1\right],
    \label{eq:f_NRH}
\end{equation}
for $\omega<\omega_d$. Finally, for the pair creation of Figure \ref{fig:cooling_EM_heat} we have a number of photons
per unit time given by:
\begin{equation}
    f'_\text{NRH}(\omega) = \frac{\pi}{4} I_B(\omega) I_B(\omega_d -\omega)
    |A_{-1}(\omega)|^2
    \label{eq:f'_NRH}
\end{equation}

This last contribution, in contrast with the other two, is not spectrally
narrow but very broad. The power spectrum associated with
the photon pairs is plotted in
Figure \ref{fig:power_spectrum}, where we can
see that these photons are
symmetrically distributed around $\omega_d/2$.
The total power spectrum if plotted in Figure \ref{fig:power_spectrum_full} for the
parameters $\omega_m/\omega_0 = .1$,
$\gamma/\omega_0 = 10^{-2}$, and $\omega_d
= \omega_0 - \omega_m$ (since this is optimal for
$\gamma/\omega_m \ll 1$).
Also, for this plot, the Dirac delta in $I_A(\omega)$ was replaced by a
Lorentzian function $(\Gamma_m/(2\pi))/((\omega-\omega_0)^2 + (\Gamma_m/2)^2)$ with a
width which was chosen as $\Gamma_m=10^{-2} \omega_m$.
When the motional mode reaches its minimal
temperature, the two main spectral peaks located at
frequencies $\omega_d\pm\omega_m$
have the same height (since the number of photons
emitted during cooling transitions should be
the same as the number of photons emitted during
heating transitions). We stress that our derivation only
rigorously applies to a system made out of harmonic
modes so the relation between our results and those
obtained for the actual model for laser cooling could be
questionable.

However, we can try to estimate the
ratio $R$ between the total number of Casimir photons and
the number of photons associated with the main peaks
when the parameters are of the same order of magnitude
than the ones typically involved in the Doppler cooling
of a trapped ion. For this, we consider the spectral density
of $\cale_B$ to be such
that $I_B(\omega) = \tilde I_B (\omega/\omega_0)^3$ (as
it is the case for the electromagnetic modes
in open space). As it is the case in the Doppler cooling
limit, we also consider that
$\omega_0 \gg \gamma \gg \omega_m$. Thus, we obtain:
\begin{equation}
  R = \frac{\int f'_\text{NRH}(\omega) d\omega}{\int f_\text{NRH}(\omega) d\omega}
  \simeq \frac{1}{4}
  \frac{\omega_m}{\omega_0} \frac{\tilde I_B}{\tilde I_A} \gamma.
\end{equation}
To pursue, we need to find an expression for the ratio
between the constants appearing in the spectral densities
of $\cale_A$ and $\cale_B$ (i.e. an expression for
$\tilde I_B/\tilde I_A$).
This can be done by choosing the parameters of our
model in such a way that they mimic the ones
corresponding to the Hamiltonian of a single trapped ion.
Then, it is simple to see that we should choose
$\tilde I_A\propto\eta^2\Omega^2$
and  $\tilde I_B\propto\gamma$,
where $\Omega$ is the Rabi frequency (which
is proportional to square root of
the laser power) and $\eta$ is the Lamb-Dicke parameter
(the ratio between the spread of the ground
state wave function and the laser wavelength).
Then, we find that
\beq
\frac{\tilde I_B}{\tilde I_A} =
\frac{\gamma}{\Omega^2 \eta^2},\nonumber
\eeq
For the typical cooling process of
Calcium ions, the experimental parameters are
\cite{roos1999}:
$\omega_m = 2\pi \times 5\text{Mhz}$,
$\omega_0 = 2\pi \times 755 \text{Thz}$ (corresponding to the $397\text{nm}$
$S_{1/2} \to P_{1/2}$ transition of a calcium ion),
$\gamma = 2\pi \times 20 \text{Mhz}$,
$\Omega = 2\pi \times 1\text{Mhz}$ and
$\eta = 0.078$. Using these values
we obtain that
$R\simeq 10^{-4}$. Also, under these conditions
approximately $3\times 10^3$ photons
per second are emitted in the cooling and heating
transitions (which, in this case, do not form two
peaks in the emission spectrum but lie inside
a single line whose width is proportional to $\gamma$).
Therefore, the number
of created pairs is exceedingly low:
less than 1 pair of photons per second are produced (in all directions),
which  makes
their observation impossible in practice (for a single trapped ion).
However, to find out if the
photon pairs created during the cooling of a
trapped ion can be observed, it is necessary to perform
a careful analysis of a realistic model (which is
out of the scope of this paper).

\section{Conclusions}

We have analyzed the solution of a simple mechanical
model which, when studied in the quantum (low temperature)
regime, can be use to understand and illustrate the most
important aspects of the process of laser cooling.
Analyzing in detail the thermodynamical properties of
linear thermal refrigerators (parametrically driven) we
estimated the lowest achievable temperatures (and the
optimal driving frequency) for the most relevant limiting
cases: the resolved sideband regime and the Doppler
(non resolved sideband) limit. The estimated temperatures
coincide with the usual expressions available in the
laser cooling literature \cite{eschner2003}: $\bar n=(\gamma/\omega_m)^2$ when $\gamma\ll\omega_m$ and
$\bar n=\gamma/\omega_m$ when $\gamma\gg\omega_m$.
The virtue of our analysis in our opinion, is that it
enables us to understand the process that fixes the
lowest achievable temperature in a simple thermodynamical
way: there is a fundamental heating process which
dominates any driven refrigerator at sufficiently low
temperatures and consists of non resonant creation of
excitations in the low frequency part of the
environmental spectrum (frequencies lower
than the driving frequency). Creation of phonon-photon
pairs unavoidably heat the motional degree
of freedom which is being cooled. It has a
counterpart in the photon-photon pairs which are
dumped in the electromagnetic environment in a way
which is entirely analogous to the dynamical Casimir
effect.  As we showed, the usual limiting temperatures
for laser cooling are obtained provided we focus in the
optical regime (where both the driving frequency and
the carrier $\omega_0$ are much larger than the motional
frequency which is being cooled). In a different
regime, our analysis enables to predict other limits, where
the lowest temperature is modified by the spectral properties
of the environment where the entropy is dumped.

We acknowledge support of ANPCyT, CONICET and UBACyT
(Argentina).

\bibliographystyle{unsrt}
\bibliography{references.bib}

\end{document}